\documentclass[conference,letterpaper]{IEEEtran}
\usepackage[utf8]{inputenc} 
\usepackage[T1]{fontenc}
\usepackage{url}
\usepackage{ifthen}
\usepackage{cite}
\usepackage[cmex10]{amsmath}
\usepackage{graphicx}
\usepackage{amsfonts}
\usepackage{enumitem}
\usepackage{enumerate}
\usepackage{amssymb}
\usepackage{booktabs}
\usepackage[boxed,ruled,commentsnumbered]{algorithm2e}

\usepackage{mathrsfs}
\usepackage{bm}
\usepackage{mdwmath}
\usepackage{subfigure}
\usepackage{mdwtab}
\usepackage{dsfont}

\newenvironment{iarray}{\begin{IEEEeqnarray}{rCl}}{\end{IEEEeqnarray}\ignorespacesafterend}

\newcommand{\tabincell}[2]{\begin{tabular}{@{}#1@{}}#2\end{tabular}}

\begin{document}
    \title{Deep Reinforcement Learning-Based Beam Tracking for Low-Latency Services in Vehicular Networks}
    \author{\IEEEauthorblockN{Yan Liu, Zhiyuan Jiang, Shunqing Zhang, and Shugong Xu,~\emph{Fellow, IEEE}}
    	\IEEEauthorblockA{Shanghai Institute for Advanced Communication and Data Science, Shanghai University, Shanghai 200444, China\\
    		Emails: LiuYann0325@163.com, \{jiangzhiyuan, shunqing, shugong\}@shu.edu.cn}    	
    	}

    \maketitle
    
    \begin{abstract}
    Ultra-Reliable and Low-Latency Communications (URLLC) services in vehicular networks on millimeter-wave bands present a significant challenge, considering the necessity of constantly adjusting the beam directions. Conventional methods are mostly based on classical control theory, e.g., Kalman filter and its variations, which mainly deal with stationary scenarios. Therefore, severe application limitations exist, especially with complicated, dynamic Vehicle-to-Everything (V2X) channels. This paper gives a thorough study of this subject, by first modifying the classical approaches, e.g., Extended Kalman Filter (EKF) and Particle Filter (PF), for non-stationary scenarios, and then proposing a Reinforcement Learning (RL)-based approach that can achieve the URLLC requirements in a typical intersection scenario. Simulation results based on a commercial ray-tracing simulator show that enhanced EKF and PF methods achieve packet delay more than $10$ ms, whereas the proposed deep RL-based method can reduce the latency to about $6$ ms, by extracting context information from the training data.
    \end{abstract}
    
    \section{Introduction}
    \label{sec_intro}    
    One of the most challenging targets for 5G and beyond cellular systems is to provide Ultra-Reliable and Low-Latency Communications (URLLC) with time-fluctuating, unreliable wireless channels. URLLC is motivated by the shifted focus of 5G systems from human-based content communications, which are relatively delay-tolerant due to limited perception capabilities of human, to machine-based control/steering information communications \cite{fet14}. It is envisioned that URLLC will enable real-time control applications in future Internet-of-Things (IoT) systems such as high-level autonomous driving, factory automation, and smart city. Despite its high expectations, URLLC in wireless networks still faces significant challenges due to e.g., pathloss, large/small-scale fading and interference. In particular, in future mm-wave-based wireless systems, wherein high beamforming gain is necessary to combat the large propagation loss of mm-wave signals, prohibitive high beam sweeping (during initial access) and tracking (while connected) overhead is entailed which becomes a severe issue for URLLC. On one hand, beam tracking/sweeping ensures good beamforming performance which is essential for high packet reception reliability; on the other hand, the incurred latency hinders URLLC---such a dilemma manifests itself in high-mobility scenarios, e.g., Vehicle-to-Everything (V2X) networks, wherein beamforming weights have to be constantly and frequently calibrated to avoid channel aging \cite{deng19}.
    
    There have been extensive works on beam alignment (including sweeping and tracking) in mm-wave channels \cite{va16,jiahui17,yavuz18,larew18,kang18,lim19,va19,chen18_globalsip,jiang19_mag}, which mainly adopt classical control theory, e.g., kalman filter-based mechanisms. In contrast, this paper presents a model-free deep Reinforcement Learning (RL) based approach. Despite the recent surge of deep learning applications in wireless network optimization, we would like to first discuss the necessity of using the learning-based approach in mm-wave URLLC systems and hence highlight two following aspects that justify its usage. First, the conventional methods which are mainly based on first principles (e.g., physical state transitions) cannot fulfill the need of URLLC, as will be illustrated in detail in the simulations. Therefore, data-driven approaches are needed since they are in nature more powerful because they can extract useful, scenario-dependent information from data, e.g., trajectory patterns, although at the expense of having to collect a large amount of data. Secondly, conventional control theory based methods mainly apply in stationary channels -- when terminals such as vehicles travel at high and time-varying speeds, the experienced wireless channel is non-stationary, thus posing severe challenges for accurate beam tracking based on Kalman filter-like schemes. Meanwhile, model-free, data-driven methods are immune to the non-stationarity. The main contributions of this paper include:
    
    1) We investigate thoroughly the state-of-the-art beam tracking methods. Due to the difficulty in obtaining the state transition equations and prior information, Extended Kalman Filter (EKF) and Particle Filter (PF) cannot be applied to non-stationary channels. Therefore, we specifically modify the EKF-based and PF-based methods for non-stationary channels. It is found that average packet delivery latency in non-stationary channels can be improved to $15$ ms and $10$ ms respectively.
    
    2) A Deep Deterministic Policy Gradient (DDPG) \cite{ddpg} based approach is proposed which extracts information and hence achieves the URLLC requirements in typical V2X networks, e.g., intersection. The evaluation is based on realistic mm-wave channels at $28$ GHz that are generated by a commercial ray-tracing simulator. It is shown that the conventional EKF- and PF-based approaches performance in non-stationary channels are not satisfactory in terms of average packet delivery latency, on account of channel training overhead and transmission failures, whereas the deep RL-based approach can reduce the delay to about $6$ ms.


\subsection{Related Work}  
     EKF is used to deal with control in nonlinear systems, and only channel gain and angle information are required in \cite{va16}. However, the state space design proposed in \cite{va16} is not applicable in non-stationary scenarios---we will modify its design and improve its performance in this paper. In \cite{jiahui17}, stochastic approximation and recursive estimation of a control parameter are used to design an algorithm that is more suitable for high-speed terminals. Ref. \cite{yavuz18} shows that Least Mean Square (LMS) performs better than EKF. However, vanilla EKF and LMS cannot work in a non-stationary scenario. The Unscented Kalman Filter (UKF), PF and Auxiliary Particle Filter (APF) are all proposed to solve the control problem under highly nonlinear systems \cite{larew18,kang18,lim19}, whose common drawback is high complexity. Beam training is carried out by an online learning algorithm in \cite{va19}, which combines online learning algorithm and hierarchical beam sweeping to select and refine beam pairs simultaneously. Ref. \cite{chen18_globalsip,jiang19_mag} apply supervised learning in beam tracking, whereas requiring labeled data. 
    \section{System Model and Problem Formulation}
    \label{sec_sm}
    In this section, we present a specific application scenario for V2X and the adopted mm-wave channel model.
    \subsection{V2X Intersection Scenario}
    \label{subsect:sys_model}
\begin{figure}[t]
		\centering    
		{\includegraphics[width=0.48\textwidth]{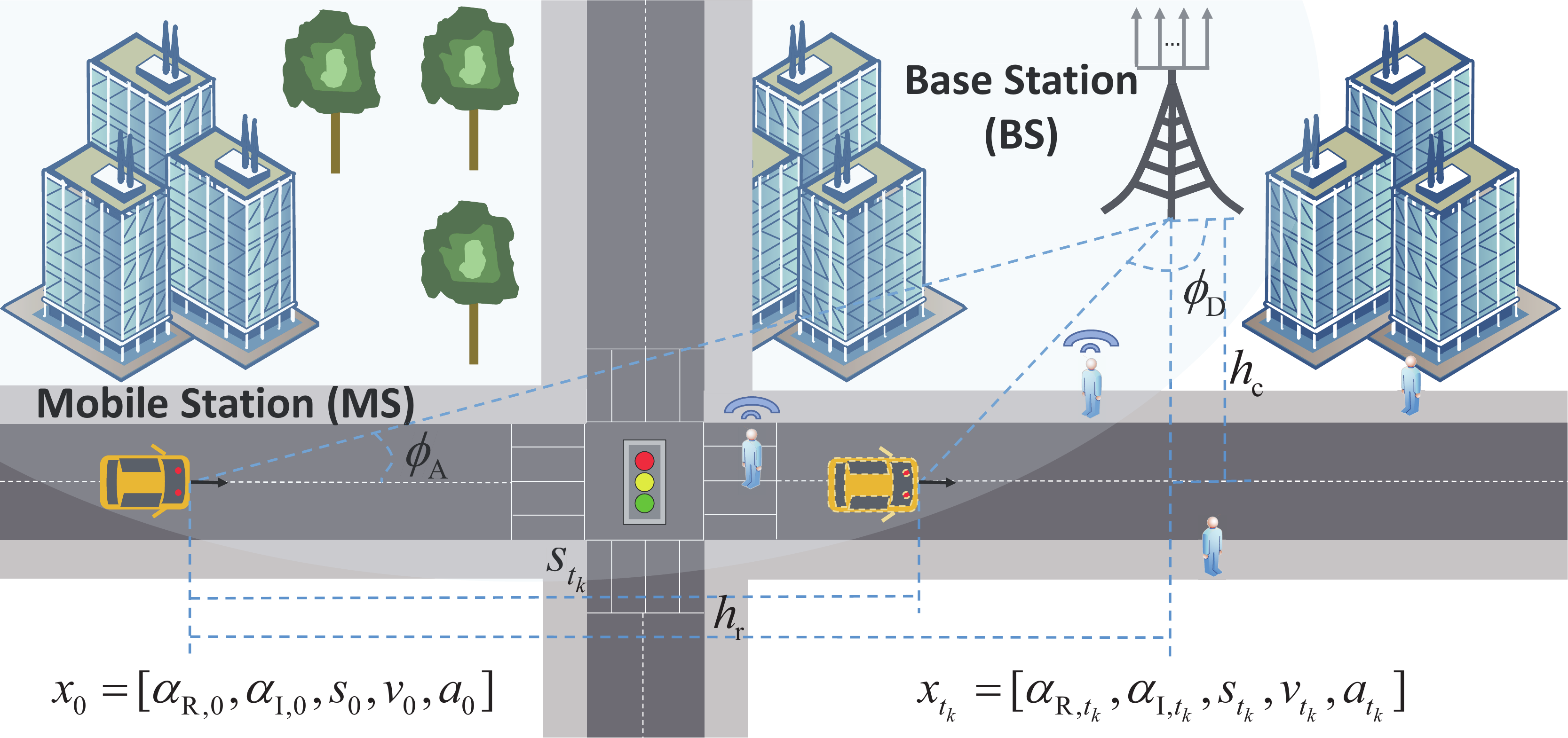}}
		\caption{The system model. We consider a typical V2X scenario in which the whole process is the movement of MS passing through the crossroad.}
		\label{fig_phy}
	\end{figure}

     We describe a typical V2X scenario with a Base Station (BS), or roadside unit, which is shown in Fig. \ref{fig_phy}. The acceleration $a_{t_k}$, velocity $v_{t_k}$, and considered moving distance $s_{t_k}$ of the Mobile Station (MS) are time-varying. The MS moves on the road. The vertical distance between the BS and the road is $h_{c}$, and $h_{r}$ represents the distance between the initial position of the MS and the vertical point. When MS is at high speed, the channel between BS and MS is non-stationary, and when the MS is waiting for traffic lights at the intersection, the MS speed is time-varying, therefore the state of the channel will change in real-time.
    \subsection{Millimeter-Wave Channel Model}
    \label{subsect:channel_model}
    We consider a Multiple Input Multiple Output (MIMO) system with Uniform Linear Arrays (ULAs). The Angle of Arrival/Departure (AoA/AoD) of the BS and MS in the Line of Sight (LoS) links are $\phi_A$ and $\phi_D$ respectively. 
    Consider a ULA of $M$ antennas, for which the array steering vector is
    \begin{align} 
        \boldsymbol{a}(\phi)=\frac{1}{\sqrt{M}}\left[1,e^{j\frac{2\pi}{\lambda}d\cos\phi},\ldots, e^{j\frac{2\pi}{\lambda}d(M-1)\cos\phi}\right]^{T},
    \end{align}
    where $\lambda$ is the carrier wavelength, $d=\frac{\lambda}{2}$ is the distance between adjacent antenna elements. We consider a time-slotted system wherein the time duration of a slot is $\Delta_{\mathrm t}$, and hence the $k$-th time slot $t_k = t_0 + k\Delta_{\mathrm t}$. The time-varying channel at the time $t_k$ can be modeled as
    \begin{equation} 
        \boldsymbol{H}_{t_k}= \sum_{l=1}^{L}\alpha_{l,{t_k}}\boldsymbol{a}(\phi_{\mathrm{A},l,{t_k}})\boldsymbol{a}^{H}(\phi_{\mathrm{D},l,{t_k}}), 
    \end{equation}
    where $L$ is the number of multi-path components, $\alpha_{l,{t_k}}$ is the channel gain of the $k$-th time slot and $l$-th path. The mm-wave channel is commonly assumed to be sparse \cite{heath16}. This paper assumes that sparsity makes the paths separate from each other and only one path falls into the main beam direction. The received signals passed by the beamformer $\boldsymbol{f}$ and the combiner $\boldsymbol{w}$ can be expressed as
    \begin{iarray} 
        y_{t_k}& = &\alpha_{i,{t_k}}\boldsymbol{w}^{H}\boldsymbol{a}(\phi_{\mathrm{A},i,{t_k}})\boldsymbol{a}^{H}(\phi_{\mathrm{D},i,{t_k}})f \nonumber\\ 
        && + \sum_{n\neq i}\alpha_{n,{t_k}}\boldsymbol{w}^{H}\boldsymbol{a}(\phi_{\mathrm{A},n,{t_k}})\boldsymbol{a}^{H}(\phi_{\mathrm{D},n,{t_k}})f+\nu_{t_k}  \nonumber\\ 
        && =\alpha_{i,{t_k}}\boldsymbol{w}^{H}\boldsymbol{a}(\phi_{\mathrm{A},i,{t_k}})\boldsymbol{a}^{H}(\phi_{\mathrm{D},i,{t_k}})f+v_{t_k}. \label{y_recieve}
    \end{iarray}
    According to \cite{mumtaz16}, the combiner and beamformer are designed to align with the beam to the direction with the maximum gain. Let $\bar{\phi}$ be the pointing direction, and the form of the beamformer/combiner can be written as $\boldsymbol{a}(\bar{\phi})=\frac{1}{\sqrt{M}}\left[1,e^{j\frac{2\pi}{\lambda}d\cos\bar{\phi}},\ldots, e^{j\frac{2\pi}{\lambda}d(M-1)\cos\bar{\phi}}\right]^{T} $. Therefore, \eqref{y_recieve} can be simplified by the geometric series formula as
    \begin{iarray}
        y_{t_k} & = &\frac{\alpha_{t_k}}{N_{r}N_{t}}\cdot\frac{1-e^{\mathrm{j}N_{r}{t_k}d(\cos\phi_{\mathrm{A}}+\cos\bar{\phi}_{\mathrm{A}})}}{1-e^{\mathrm{j}{t_k}d(\cos\phi_{\mathrm{A}}+\cos\bar{\phi}_{\mathrm{A}})}} \nonumber\\
         && \cdot\frac{1-e^{\mathrm{j}N_{t}{t_k}d(\cos\phi_{\mathrm{D}}+\cos\bar{\phi}_{\mathrm{D}})}}{1-e^{\mathrm{j}{t_k}d(\cos\phi_{\mathrm{D}}+\cos\bar{\phi}_{\mathrm{D}})}} +v_{t_k},  \label{equa_y}
    \end{iarray}
    where $N_\textrm{t}$ is the number of antennas at the transmitter and $N_\textrm{r}$ at the receiver.
    \subsection{Problem Formulation}
    Our goal is to reduce the average packet delivery latency during the time a vehicle moves through the intersection with guaranteed reliable performance to meet the low-latency requirements in URLLC. In the learning process, the position and motion of MS over time are unknown to BS, hence it is possible that the beam is not aligned with the direction of the MS. Furthermore, whether a packet can be transmitted successfully depends on the accuracy of beam alignment, and both beam tracking and retransmissions are time-consuming. The tradeoff between beam alignment and data transmission should be balanced to obtain the minimum delay and the maximum number of successful delivery packets. 
    
    \section{The Proposed DDPG-Based Algorithm}
    \label{sec_mr}
    
    \begin{figure}[t]
		\centering    
		{\includegraphics[width=0.48\textwidth]{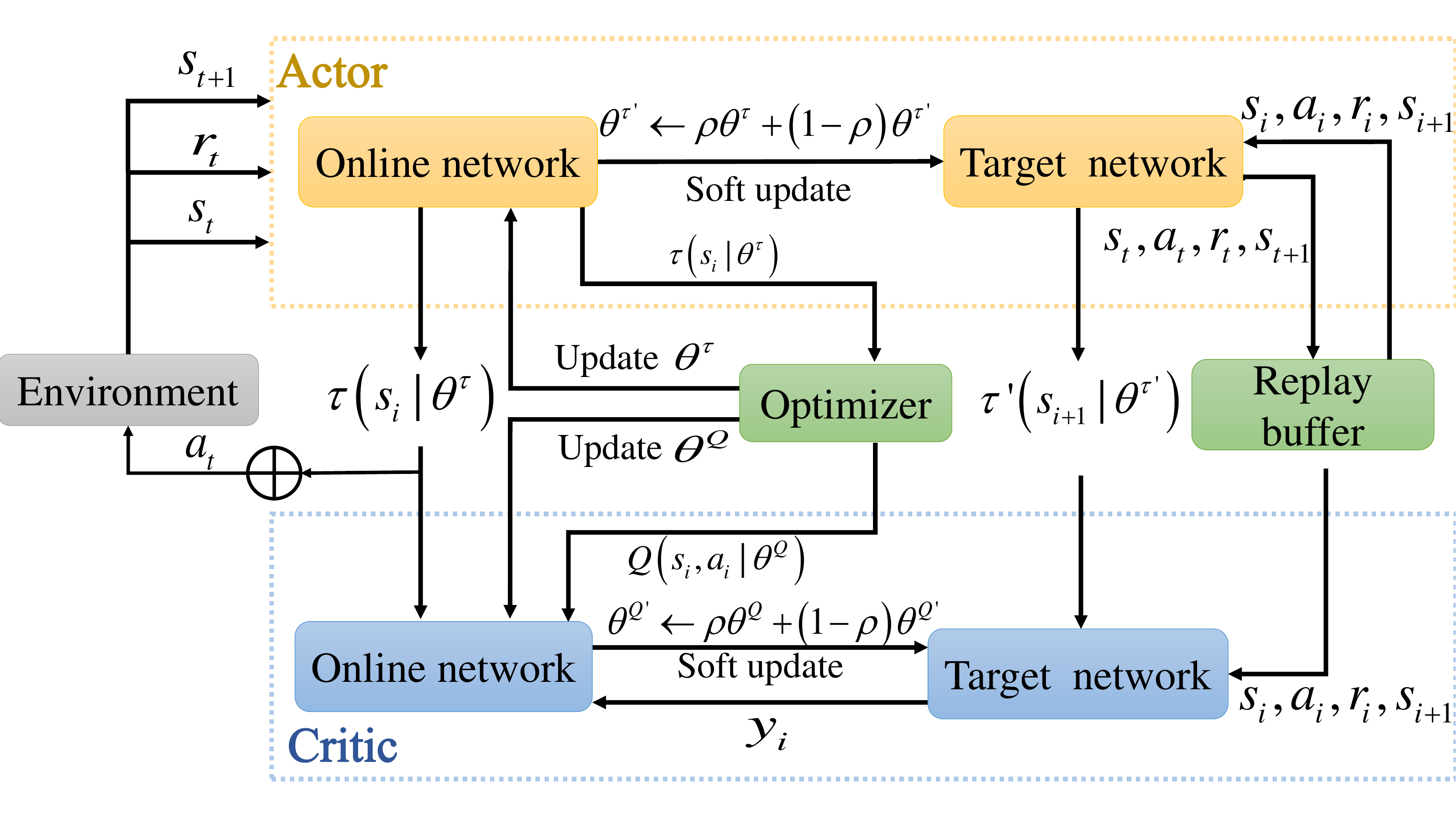}}
		\caption{The overall structure of DDPG. This includes actor network and critic network, both of which contain online network and target network respectively. The important data flow is shown in the figure.}
		\label{fig_ddpg}
	\end{figure}
    The DDPG algorithm is based on model-free and off-policy RL techniques, meanwhile, a deep neural network is used for function approximations. Different from traditional algorithms, DDPG can solve problems with continuous action space. In Fig. \ref{fig_ddpg}, DDPG consists of two networks. A neural network to approximate the value function. This value function network is also called the critic network, whose input is action and observation, the output is a value of the state-action pair, i.e., $Q(s,a)$; In addition, a neural network is used to approximate the policy function, which is also known as the actor network. Its input is observation value and output is action value. We use $\theta^Q$ and $\theta^\tau$ to parameterize function approximators. Hence, we design the DDPG-based algorithm which explores the tradeoff between beam tracking overhead and data transmission and achieves the minimal packet delivery latency.
    
    The flow of data between different networks is shown in Fig. \ref{fig_ddpg},  actor network obtains $s_t$ from the environment. After data flow, action $a_t$ act on the environment to get $r_t$, and then obtains $s_{t+1}$ from the environment. In the DDPG algorithm, the critic network is updated by minimizing the loss:
    \begin{equation}
        L(\theta^Q)= \frac{1}{R} \sum_i(y_i-Q(s_{i}, a_{i}\vert\theta^{Q}))^{2}, \label{critic1}
    \end{equation}
    where $y_i = r_i + \gamma Q^{\prime}(s_{i+1},\tau^{\prime}(s_{i+1}\vert{\theta^{\tau^{\prime}}})\vert{\theta^{Q}})^{2},\label{critic2}$ and $\gamma$ is the discount factor. At the same time, we optimize the actor network by maximizing the policy objective function $J$:
    \begin{equation}
        \nabla_{\theta^{\tau}}J\approx\frac{1}{R}\sum_i\nabla_{a}Q(s,a\vert\theta^{Q})\vert_{s=s_i,a=\tau(s_i)}\nabla_{\theta^{\tau}}\tau(s\vert \theta^{\tau})\vert _{s_i}. \label{actor}
    \end{equation}
    
    The details of the proposed DDPG-based algorithm are described in Alg. \ref{alg1}. The state space of the algorithm is defined as $\mathcal{S} = \{\omega,y_R,y_I,T\}$, where $\omega$ is the beam angle of the current time-slot, $y_R$ and $y_I$ are the real and imaginary components of the observed signal \eqref{equa_y}, respectively, and $T$ is the time interval between the last beam tracking time and the current time. The action space is continuous and two-dimensional, which is denoted by $\mathcal{A} = \{a_b,a_f\}$. The former controls beam direction, and the latter takes charge of whether at the current time step, the system performs beam direction correction. The delay of the sending packet is represented by a reward $\mathcal{R}$. Specifically, if a packet is transmitted successfully, the delay remains unchanged. If not, the delay is increased (reward is decreased) by a time slot.  An episode is one run of an MS at the intersection. The agent in DDPG is, in this case, the BS that interacts with the intersection environment through a period of observations, actions, and rewards to optimize the average delay of data transmission in the process of MS movement. In the training phase of DDPG, in order to reduce training overhead and decision time, a step contains several time slots. In addition, the agent (BS) determines whether the current step needs beam tracking. If the decision is to transmit data, all time slots of the current step are used to send data, and then the number of packets and total delays are counted. If the beam direction needs to be calibrated, the first time slot of the step is used for beam tracking, and the remaining time slots are used for data transmission. 
    
    \begin{algorithm}[t]
    	\caption{DDPG-based beam tracking and data transmission algorithm}
    	\label{alg1}
    		\label{alg:init}
    		Initialization: \\
    		The critic network $Q(s,a\vert{\theta^{Q}})$ and actor network $\tau(s\vert{\theta^{\tau}})$ with weights $\theta^{Q}$ and $\theta^{\tau}$;\\
            The target network $Q^{\prime}$ and $\tau^{\prime}$ with weights $\theta^{Q^{\prime}}\gets{\theta^{Q}}$ and $\theta^{\tau^{\prime}}\gets{\theta^{\tau}}$;\\
            Replay buffer $B$, actor and critic learning rate $LR_\mathrm{A}/LR_\mathrm{C}$, batch size $m$ and memory capacity $R$;\\
    		\label{alg:start}
            \For {$episode = 1 , N$}{
                Obtain initial observation state $s_1$ from environment;\\
                Reset the total number of packets ${ep}_{packet}$ and the total reward value ${ep}_{reward}$ in one episode;\\
                \While{$t \leq E$ or $done \neq terminal$}{
                Select action $a_t = \tau(s_t\vert{\theta^{\tau}})+\mathcal{N}$ according to the current policy and exploration noise;\\
                Execute action $a_t$, then observe next state $s_{t+1}$, reward $r_t$, step end or not $done$ and the number of packets in this step $n_{packet}$;\\
                Store tuple $(s_t,a_t,r_t,s_{t+1})$ in $B$;\\
                Sample a batch size $m$ of $R$ transitions $(s_i,a_i,r_i,s_{i+1})$ from $B$;\\
                Update the critic $\theta^Q$ by Eq. \eqref{critic1};\\
                Update the actor $\theta^\tau$ using Eq. \eqref{actor};\\
                Update the target networks:\\
                $\theta^{Q^{\prime}}\leftarrow \rho \theta^{Q}+(1-\rho)\theta^{Q^{\prime}};$\\ $\theta^{\tau^{\prime}}\leftarrow \rho \theta^{\tau}+(1-\rho)\theta^{\tau^{\prime}};$\\
                $t = t+1;$
                }
            }
    \end{algorithm}
    \begin{table}[!t]
	\centering
	\caption{DDPG Neural Network Composition}
	\label{tab_net}
	\begin{tabular}{lllll}
	\toprule
	\multicolumn{1}{c}{Layer} & \multicolumn{1}{c}{\tabincell{c}{Actor's \\network \\size}} & \multicolumn{1}{c}{\tabincell{c}{Actor's \\activation \\function}} & \multicolumn{1}{c}{\tabincell{c}{Critic's \\network \\size}} & \multicolumn{1}{c}{\tabincell{c}{Critic's \\activation \\function}}\\
	\hline
	\multicolumn{1}{c}{Input} & \multicolumn{1}{c}{$D_\textrm{state}$} & \multicolumn{1}{c}{\tabincell{c}{ReLU}} & \multicolumn{1}{c}{$D_\textrm{state}+D_\textrm{action}$} & \multicolumn{1}{c}{ReLU}\\
	\hline
	\multicolumn{1}{c}{Hidden layer1} & \multicolumn{1}{c}{$N_i$} & \multicolumn{1}{c}{\tabincell{c}{ReLU}} & \multicolumn{1}{c}{$N_i$} & \multicolumn{1}{c}{ReLU}\\
	\hline
	\multicolumn{1}{c}{Hidden layer2} & \multicolumn{1}{c}{$N_i$} & \multicolumn{1}{c}{\tabincell{c}{ReLU}} & \multicolumn{1}{c}{$1/20N_i$} & \multicolumn{1}{c}{ReLU}\\
	\hline
	\multicolumn{1}{c}{Hidden layer3} & \multicolumn{1}{c}{$1/20N_i$} & \multicolumn{1}{c}{\tabincell{c}{ReLU}} & \multicolumn{1}{c}{/} & \multicolumn{1}{c}{/}\\
	\hline
	\multicolumn{1}{c}{Output} & \multicolumn{1}{c}{$D_\textrm{action}$} & \multicolumn{1}{c}{ReLU} & \multicolumn{1}{c}{1} & \multicolumn{1}{c}{ReLU}\\
	\bottomrule
	\end{tabular}
	\end{table}

	Finally, we analyze the network architecture of DDPG in Tab. \ref{tab_net}, where $N_i$ is the number of neurons in each layer. $D_\textrm{state}$ and $D_\textrm{action}$ are the dimensions of state and action space, respectively. The actor network contains three fully-connected hidden layers and the critic network contains two fully-connected hidden layers. Through the interaction of the two networks, we can know the relationship between the vehicle state and the beam state. 
    \section{Comparative Schemes and Simulation Results}
    In this section, in order to obtain a thorough understanding of the state-of-the-art and its limitations in non-stationary channels, we first describe and improve two well-known beam tracking schemes, namely EKF and PF schemes, and then present simulation results comparing these two with our proposed scheme.
    \begin{figure}[!t]
		\centering    
		{\includegraphics[width=0.48\textwidth]{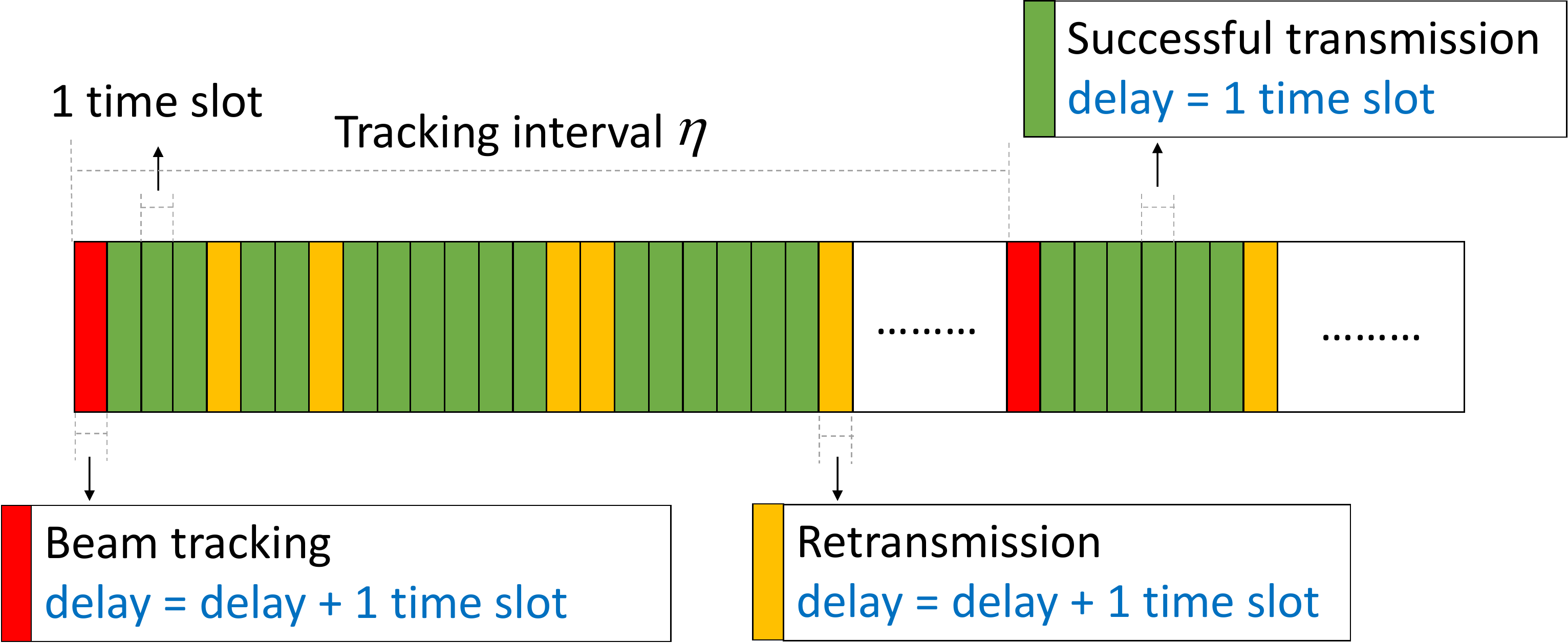}}
		\caption{Explanation of the frame structure used in EKF- and PF-based schemes. Red, green and yellow blocks represent beam tracking, successful transmission and unsuccessful transmissions that require retransmissions, respectively. When beam tracking and retransmissions are carried out, the total delay of the packet is added by one time slot, whereas the total delay remains unchanged when the transmission is successful. Finally, the average transmission delay of the packet is given by the ratio between the total delay and the total number of successfully transmitted packets.}
		\label{fig_fram}
	\end{figure}
    \subsection{Relevant Comparative Schemes}
    \begin{figure*}[t]
		\centering    
		{\includegraphics[width=0.7\linewidth]{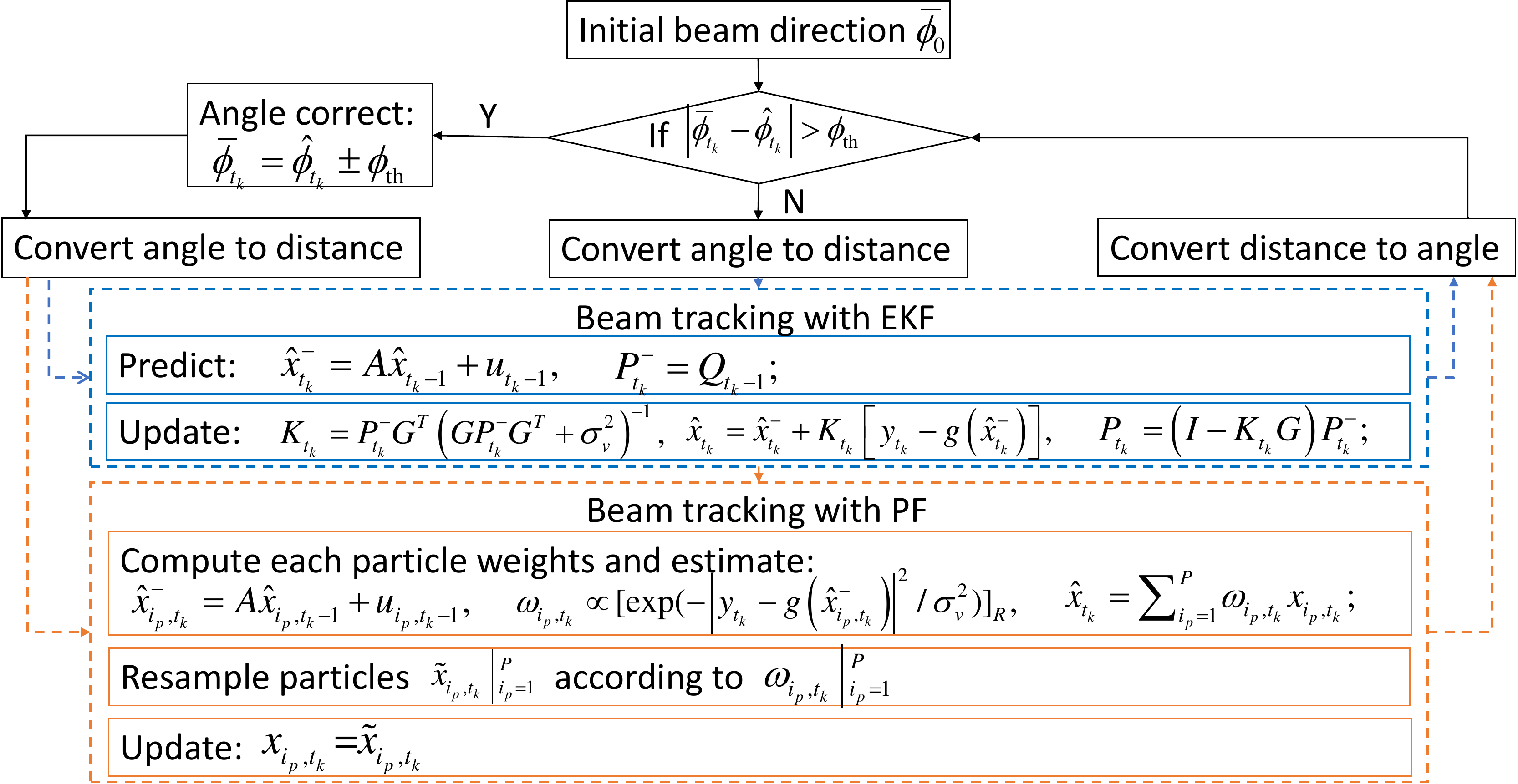}}
		\caption{An overview of EKF-based and PF-based schemes. In order to better adapt to the non-stationary scenario, the variables in state space change from angle to motion distance, velocity and acceleration.}
		\label{fig_ekf}
	\end{figure*}
    First, the frame structure of the transmitted signal is explained for EKF and PF schemes. The transmission is divided into two phases: 1) beam tracking and 2) data transmission. As shown in Fig. \ref{fig_fram}, pilot symbols are sent in phase 1. In phase 2, data transmission is performed, and the delay of each packet is calculated based on the total number of packets transmitted and the total delay. 
    
    Then, an overview of the beam tracking process by EKF and PF is shown in Fig. \ref{fig_ekf}. Since this paper focuses on beam tracking and data transmission process, the angle estimation is not considered. When the difference between the predicted direction and the pointing direction exceeds the threshold $\phi_\mathrm{th}$, the correction is carried out according to the method in \cite{va16}. Note that the original EKF and PF schemes cannot deal with non-stationary channels. In particular, the state transition matrix is not obtainable, and in fact, if we simulate the vanilla EKF and PF schemes, their performance is so poor that it cannot be compared with our proposed scheme on a reasonable scale. Therefore, we apply adjustments to both schemes for proper comparison. Specifically, the state space is defined as
    \begin{equation}
        \boldsymbol{x}_{t_k}=[\alpha_{\mathrm{R},{t_k}}, \alpha_{\mathrm{I},{t_k}} , s_{t_k} , v_{t_k} ,a_{t_k}]^{\mathrm{T}},
    \end{equation}
    where $\alpha_{t_k} = \alpha_{\mathrm{R},{t_k}}+\mathrm{j}\alpha_{\mathrm{I},{t_k}}$ is the channel gain. The state transition model can be written as 
    \begin{equation}
        \boldsymbol{x}_{t_k}=\boldsymbol{A}\boldsymbol{x}_{t_{k-1}}+\boldsymbol{u}_{t_k} \label{equa_x}, 
    \end{equation}
    where 
    \begin{equation}
        \boldsymbol{A} = \begin{bmatrix}\rho & 0 & 0 & 0 & 0 \\ 0 & \rho & 0 & 0 & 0 \\ 0 & 0 & 1 & \Delta_\mathrm{t} & \frac{\Delta_{\mathrm{t}}^{2}}{2} \\ 0 & 0 & 0 & 1 & \Delta_\mathrm{t} \\ 0 & 0 & 0 & 0 & 1\end{bmatrix} \label{equa_A},
    \end{equation}
    and $\rho$ denotes the correlation coefficient and the noise is $\boldsymbol{u}_{t_k}\sim\mathcal{N}(\boldsymbol{0},\boldsymbol{Q}_{t_k})$ with $\boldsymbol{Q}_{t_k}=\text{diag}([1-\rho^{2},1-\rho^{2},\ \sigma_{u}^{2}\Delta_{\mathrm{t}}^{2}/2,\ \sigma_{u}^{2}\Delta_\mathrm{t},\ \sigma_{u}^{2}]), $ where $\sigma_{u}^{2}$ is the standard deviation.
	
	Kalman filtering is a minimum variance estimation scheme. The implementation of EKF necessitates to meet two conditions: 1) the initial state must be drawn from a normal distribution; 2) the system cannot be highly nonlinear. PF is derived from the idea of Monte-Carlo, which samples variables and approximates the distribution with a large number of samples during the filtering process. Therefore, PF can handle non-Gaussian distribution, while KF can only deal with Gaussian distribution. EKF improves upon KF, linearizing the nonlinear problem near the operating point, but the linearization process takes the first-order Taylor expansion and thus loses performance in highly-nonlinear scenarios. The complexity of EKF and PF are compared in \cite{lim19}, showing that the EKF complexity is $\mathcal{O}(D)$, and the algorithm complexity of PF is $\mathcal{O}(DP)$, where $D$ is state dimension and $P$ is the number of particles. The reason why PF is more complex than EKF is that PF needs to derive the weight of each particle from the variables of each dimension in the state space and sum them together to get the new estimation.
    
	\subsection{Performance Evaluation}
    To compare the performance of the proposed DDPG-based approach and the baseline of modified EKF and PF schemes, computer simulations are performed and results are presented in this subsection. The simulation parameters and some network hyperparameters are listed in the Tab. \ref{table1}.
    \begin{table}[!t]
	\centering 
	\caption{Simulation Parameters} 
	\label{table1}  
	\begin{tabular}{cc}  
		\hline  
		Parameter/Hyperparameter & Value \\  
		\hline
		Number of transmit and receive antennas $N_\mathrm{t}$=$N_\mathrm{r}$ & $16$ \\
		Initial beam direction & $3\pi/4$ \\
		Initial distance $s_0$ & $0$ [m]\\
		Initial acceleration $a_0$ & $-4$ [$\mathrm{m/s^2}$]\\
		Initial velocity $v_0$ & $16$ [m/s]\\
		Correlation coefficient $\rho$ & $0.995$\\
		Time slot $\Delta_\mathrm{t}$ & $5$ [ms]\\
		Total time $t_\mathrm{total}$ & $10$ [s]\\
		The vertical distance between \\the BS and the road link $h_c$ & $200$ [m]\\
		The distance between the initial position of \\the MS and the vertical point $h_r$ & $200$ [m]\\
		Max step $E$ & $1000$\\
		Max episode $N$ & $1800$\\
		Discount factor $\gamma$ & $0.9$\\
		Batch size $m$ & $16$\\
		Memory capacity $R$ & $5000$\\
		The number of particles $P$ & $1000$\\
		Actor and critic network learning rate $(LR_\mathrm{A},LR_\mathrm{C})$ & $(10^{-4},10^{-4})$\\
		Number of neurons in network layer $N_i$ & $200$ \\
		\hline
	\end{tabular}
    \end{table}
    
    For simplicity, we assume the following MS mobility process to simulate vehicle movement at the intersection: firstly it uses 4 seconds to decelerate, waits 2 seconds, then uses 4 seconds to accelerate again. The initial velocity has reached $16$ m/s. At this high speed, the channel is already non-stationary. Note that such a pattern can be adapted to real-world cases whereas the current case is for illustration of the beam tracking scheme. In order to make a fair comparison between the baseline algorithms and the proposed algorithm without making the training of the neural network too slow, a time slot of 5 ms is adopted, and the EKF and PF schemes track beams at a certain time interval, while DDPG algorithm determines beam alignment every 20 time slots. 
    
    For URLLC considerations, the target BLock Error Rate (BLER) is set to $10^{-6}$. According to \cite{sybis16}, SNR needs to be higher than about $5$ dB for a packet with typical Modulation and Coding Scheme (MCS) and Tail-Biting Convolutional Code (TBCC) to be sent successfully. First, we use channel data from MATLAB-simulated channel models to illustrate the performance comparisons, then we use ray-tracing data to test the convergence of the proposed scheme.
    \begin{figure}[!t]
		\centering    
		{\includegraphics[width=0.48\textwidth]{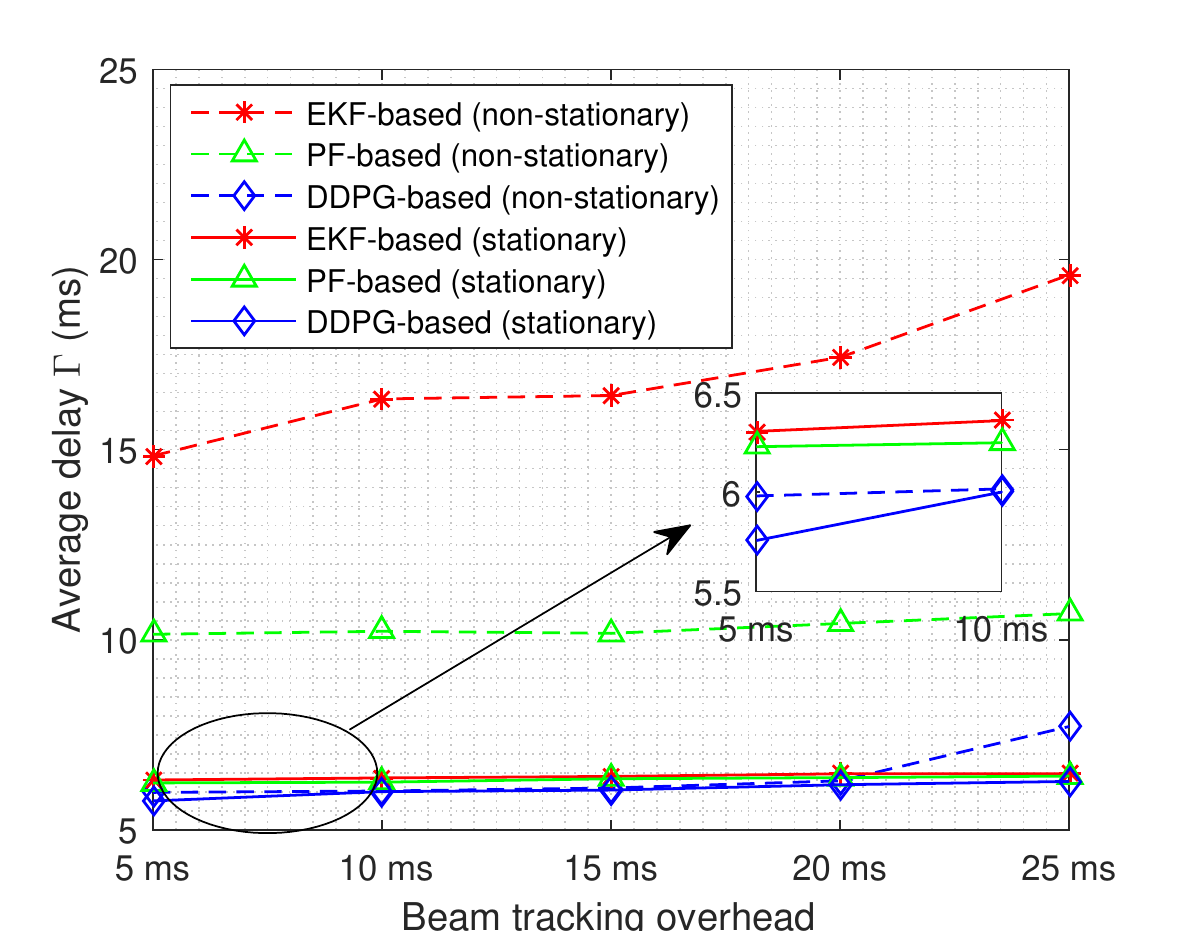}}
		\caption{Results of different beam tracking overhead in non-stationary/stationary channels. The DDPG-based algorithm can achieve the optimal average delay performance of $6$ ms by learning in either scenario. }
		\label{fig_over}
	\end{figure}
    \subsubsection{Simulation results based on LoS channel models}
\begin{figure}[!t]
		\centering    
		{\includegraphics[width=0.48\textwidth]{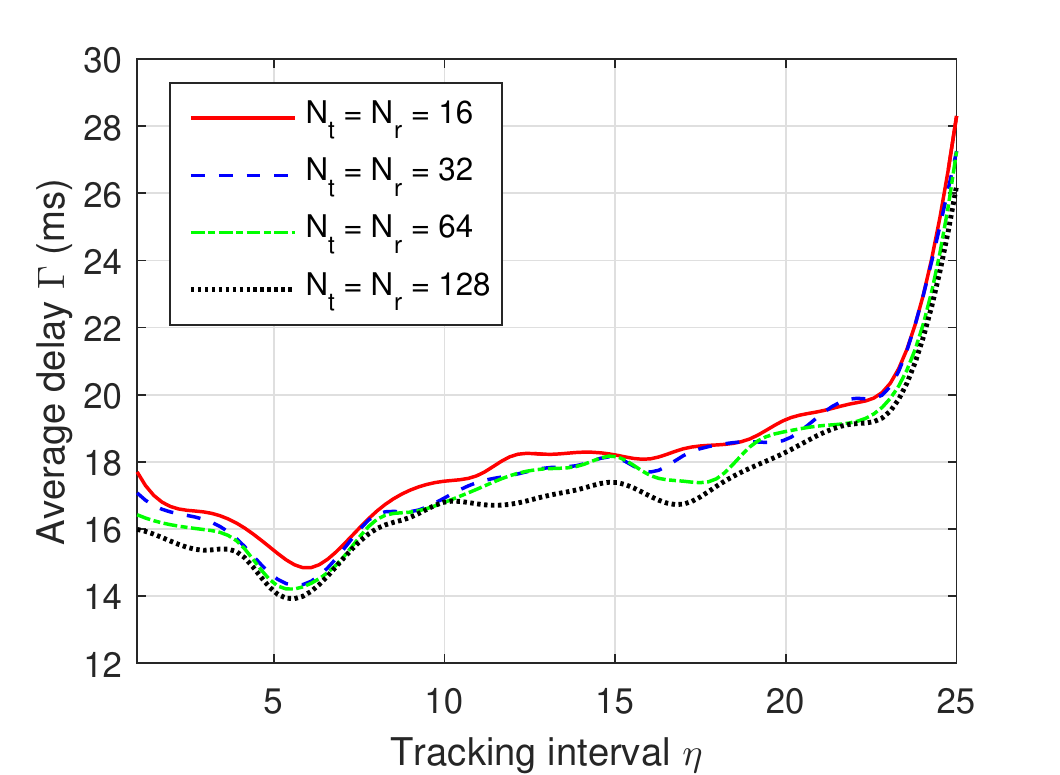}}
		\caption{Results of different beam tracking intervals and antenna array sizes using the EKF-based algorithm. The increase of the number of antennas brings a small gain for the average packet delay, and the optimal value is reached when the beam tracking interval is $0.2$ s in this case.}
		\label{fig_inter}
	\end{figure}

    We use MATLAB to model a LoS channel according to the descriptions in Sec.\ref{sec_sm}. Fig. \ref{fig_over} shows the impact of different channel training overhead, which is expressed by the average packet delivery delay of packets when the beam tracking interval is $0.1$ s. We can observe that the minimum delay of the EKF-based algorithm is about $15$ ms, the PF-based algorithm is about $10$ ms and the DDPG-based algorithm can reach about $6$ ms. In stationary channels, i.e., MS moves at a constant and small speed, three algorithms perform equally well since EKF and PF are both suitable for stationary systems as long as the state transition is estimated correctly. Meanwhile, more beam tracking overhead leads to less time for data transmission, and packet delays will increase when the number of successfully transmitted packets decreases. For stationary channels, we assume that the MS moves at a constant speed of $8$ m/s. Fig. \ref{fig_inter} compares the effects on packet transmission delay of EKF from two aspects -- the channel tracking interval and the number of antennas. We see that the average delay of packets is lowest when beam tracking is carried out every $0.2$ s. When the tracking interval is too long, a beam may not track its target, resulting in a packet with a higher BLER requiring multiple retransmissions. On the contrary, the system spends too much time on beam tracking, leading to a long overall delay and a small number of packets sent. Moreover, as the number of antennas increases, the packet transmission delay decreases slightly as we notice. This is due to large antenna arrays have a narrower beam requiring more accurate tracking. Based on Fig. \ref{fig_over} and Fig. \ref{fig_inter}, we can conclude that the DDPG-based algorithm has a tremendous improvement compared with EKF-based and PF-based algorithms in non-stationary scenarios.

\begin{figure}[!t]
    \subfigure[In LoS scenario]{
    \begin{minipage}[!t]{1\linewidth}
    \begin{center} \includegraphics[width=1\textwidth]{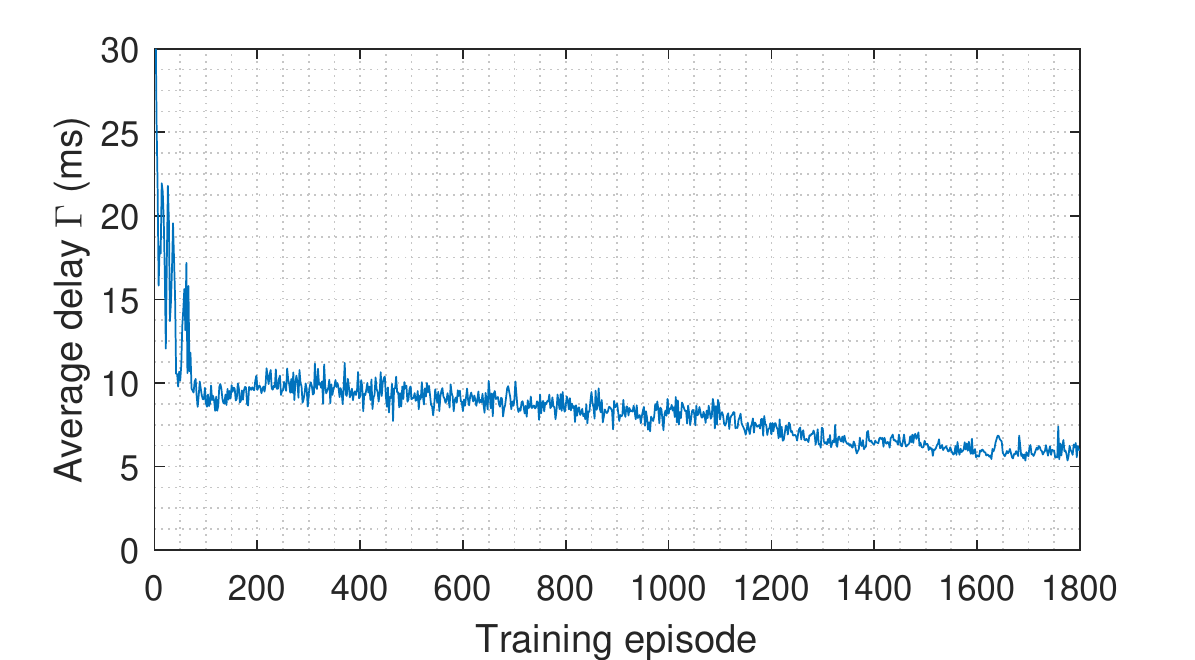}
    \end{center}
    \label{fig_rldelay}
    \end{minipage}
    }
    \subfigure[In realistic mm-wave channels]{
    \begin{minipage}[!t]{1\linewidth}
    \begin{center}
    \includegraphics[width=1\textwidth]{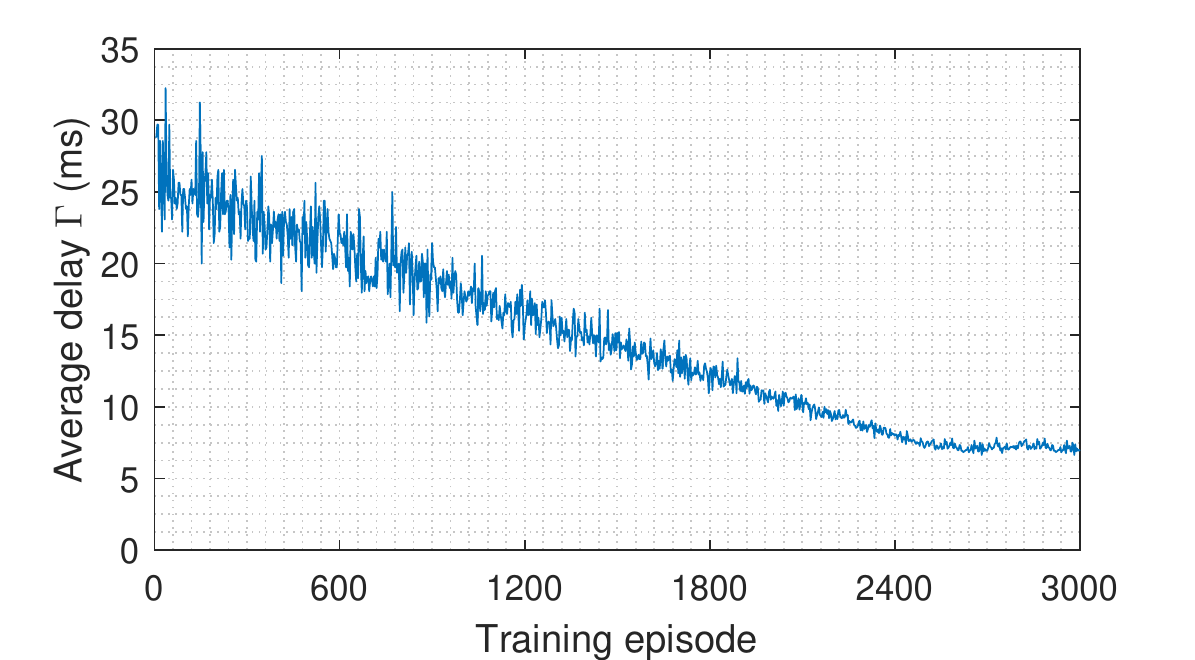}
    \label{fig_rtdelay}
    \end{center}
    \end{minipage}
    }
    \caption{Training results of the DDPG-based algorithm. $LR_\mathrm{A} = LR_\mathrm{C} = 10^{-4}, \gamma = 0.9$. As shown in Fig. \ref{fig_rldelay}, an episode contains $1000$ steps and a step represents $0.1$ s and the average transmission delay of packets gradually converges to about $6$ ms after about $100$ episodes. As shown in Fig. \ref{fig_rtdelay}, an episode contains $600$ steps and a step represents $0.05$ m of movement and the average packet delay finally converges to about $7$ ms after about $2520$ episodes.}
    \end{figure}

\begin{figure}[!t]
    \centering
    \subfigure[Discount factor $\gamma$]{
    \begin{minipage}[t]{0.45\linewidth}
    \centering
    \includegraphics[width=1\textwidth]{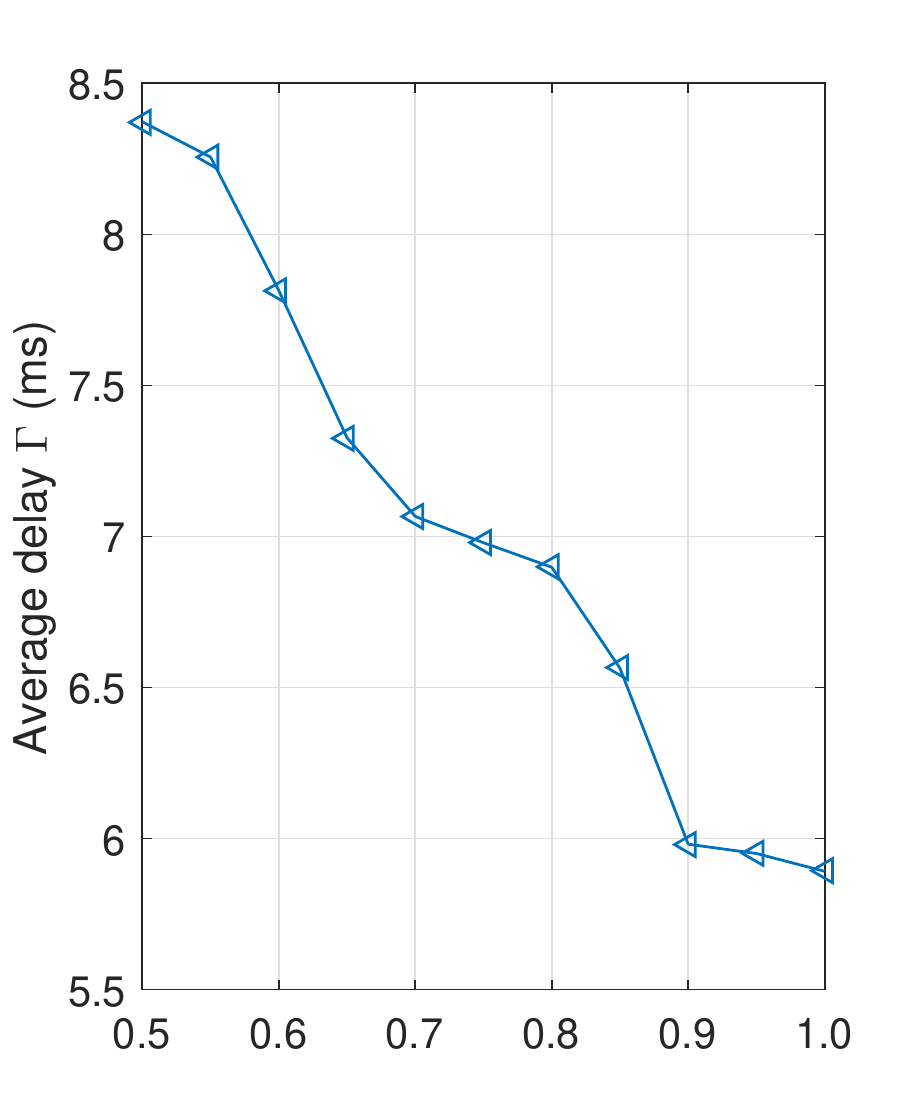}
    \label{fig_df1}
    \end{minipage}
    }
    \subfigure[Learning rate ($LR_\mathrm{A}$,$LR_\mathrm{C}$)]{
    \begin{minipage}[t]{0.45\linewidth}
    \centering
    \includegraphics[width=1\textwidth]{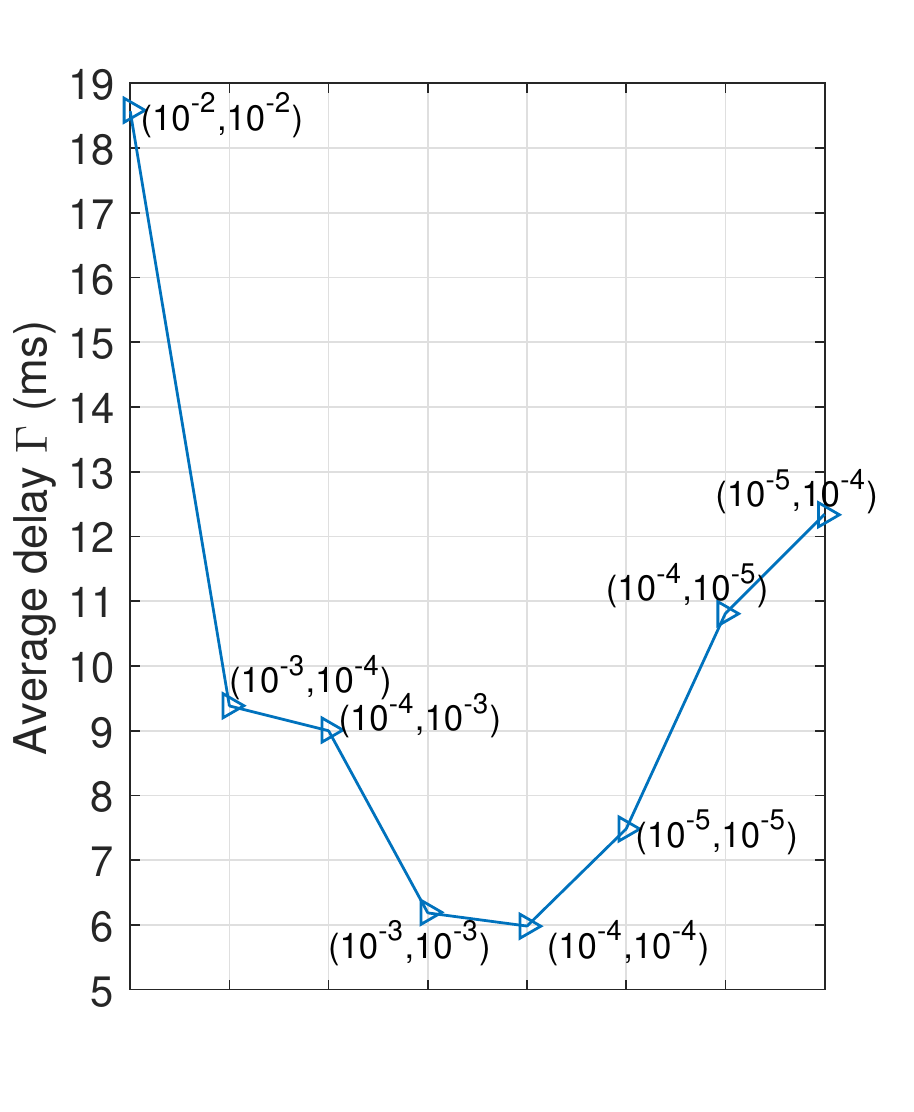}
    \label{fig_lr1}
    \end{minipage}
    }
    \centering
    \caption{Results of different discount factors and learning rates. In Fig. \ref{fig_df1}, $LR_\mathrm{A} = LR_\mathrm{C} = 10^{-4}$ and we can find that the average packet delay for the larger discount factor is higher. In Fig. \ref{fig_lr1}, $\gamma = 0.9$ and learning rates of the actor network and the critic network are respectively represented in the bracket; as can be seen $(10^{-4},10^{-4})$ outperforms others.}
    \end{figure}

    In the DDPG training process, the average packet delay over time is shown in Fig. \ref{fig_rldelay}. It is observed that DDPG converges rapidly, with a sharp decline around the $100$th episode and the average transmission delay of packet finally converges to about $6$ ms. In addition, we compare the average delay by a different discount factor $\gamma$ in Fig. \ref{fig_df1}. At last, $\gamma = 0.9$ is selected as our discount factor. It is observed from Fig. \ref{fig_lr1} that different learning rates have an effect on performance and that the loss is larger when the learning rates of the actor and critic networks are different. Consequently, we choose $LR_\mathrm{A} = LR_\mathrm{C} = 10^{-4}$ as our learning rate.
    
    
    
    \subsubsection{Simulation results based on ray-tracing channel data}
    In the second part, the simulation is extended to realistic mm-wave channel data generated by ray-tracing at $28$ GHz frequency. In most studies, channel measurement data are difficult to obtain, and ray-tracing data (obtained from Wireless Insite) are close enough to reality \cite{f15}. The considered scenario mainly involves the direct ray that is not blocked by the building or trees and reflected paths by the building and trees. The velocity, acceleration and motion distance of the MS are the same as described before. Because ray-tracing models produce channel data that have multi-path components, the conventional EKF and PF schemes are hard to converge in this case, and thus the simulation results are not shown.
    
    After using the data generated by the ray-tracing simulator to train DDPG, the results are reflected in Fig. \ref{fig_rtdelay}. In this experiment, the maximum episode is $3000$ which each contains $600$ steps. The learning rate and discount factors are the same as the previous simulation. It can be seen from the figure that the average delay based on data from ray-tracing converges relatively slowly due to multi-path components, but eventually converges to about $7$ ms, i.e., the DDPG-based approach also performs excellently in realistic mm-wave channels.
    \section{Conclusions}
    This paper studies the beam tracking problem that satisfies URLLC in mm-wave MIMO systems and proposes a DDPG-based approach for typical V2X scenarios. In addition, we improve the traditional EKF and PF methods to enable them to be applied in a non-stationary environment. Based on ray-tracing-based simulation results, in the non-stationary scenarios and stationary scenarios, the lowest average packet delay of the EKF-based algorithm can reach $15$ ms and $6.3$ ms respectively, while the lowest average packet delay of the PF-based algorithm is $10$ ms and $6.2$ ms respectively. The proposed DDPG-based scheme learns the non-stationarity of the scenario and hence achieves an average delay of packets in both scenarios as low as $6$ ms, which makes it favorable in future mm-wave URLLC scenarios.
    \bibliographystyle{ieeetr}
    \bibliography{EKF_RL}
    \end{document}